\documentclass[aps,showpacs,floatfix,twocolumn]{revtex4}
\usepackage{amsmath}
\usepackage{graphicx}
\usepackage{floatflt,epsfig}
\usepackage{lscape}
\usepackage{ulem}
\def\beq{\begin{equation}}
\def\eeq{\end{equation}}
\def\beqa{\begin{eqnarray}}
\def\eeqa{\end{eqnarray}}
\def\ban{\begin{eqnarray*}}
\def\ean{\end{eqnarray*}}
\def\bi{\begin{itemize}}
\def\ei{\end{itemize}}

\begin{document}

\title{Dynamical instabilities of warm $npe^-$ matter:\\ the $\delta$ meson effects}

\author{Helena Pais, Alexandre Santos and Constan{\c c}a Provid\^encia}
\affiliation{Centro de F\'{\i}sica Computacional, Department of Physics\\
University of Coimbra, 3004-516 Coimbra, Portugal}

\begin{abstract}
The effects of $\delta$ mesons on
the dynamical instabilities of cold and warm nuclear and stellar
matter at subsaturation densities
are studied in the framework of relativistic
mean-field hadron models (NL3, NL$\rho$ and NL$\rho\delta$) with the
inclusion of the electromagnetic field. The distillation effect and the spinodals for all the models considered are discussed.
The crust-core transition density and pressure are obtained as a
function of temperature for $\beta$-equilibrium matter with and without
neutrino trapping. An estimation of the size of the clusters formed in
the non-homogeneous phase is made. It is shown that cluster sizes increase with
temperature.
The effects of the $\delta$-meson on the
instability region are larger for
low temperatures, very asymmetric matter and  densities  close to the
spinodal surface. It increases the distillation effect above $\sim 0.4\rho_0$ and
has the opposite effect below that density.
\end{abstract}


\maketitle

\section{Introduction}

Compact stars are to date believed to be made of inner layers enclosed in a crust and possibly, a shallow atmosphere \cite{Prakash-97}. The information we can collect from the interior of the star has necessarily crossed its crust. Therefore, a  better understanding of these compact objects demands a precise description of their envelope. The inner crust of compact stars is described by a liquid-gas phase
transition of asymmetric nuclear matter in the presence of electrons. In this context, the isospin content of the liquid and the gas phases can play an important role in transport phenomena and as consequence, account for the energy losses from neutrino emissivity \cite{Sawyer-75,Iwamoto-82}.\\
Several theoretical and experimental efforts have been set in order to more precisely describe the physics involved in these phenomena, and there is currently considerable research and a number of experiments being done in the search for a better description of nuclear matter under exotic conditions (namely of density, temperature and asymmetry).

The authors of Ref. \cite{Liu-02,Greco-03} have stressed the importance of including the scalar
isovector virtual $\delta$($a_0$(980)) field in hadronic effective field theories when
asymmetric nuclear matter is studied. Its presence introduces in the isovector channel the
structure of relativistic interactions, where a balance between a scalar (attractive) and a
vector (repulsive) potential exists. The $\delta$, and $\rho$ mesons give rise to the
corresponding attractive and repulsive potentials in the isovector channel. The introduction of
$\delta$ mesons  affects the behavior of the system at high densities, when its contribution
is reduced  leading to a harder equation of state (EOS) and at 
subsaturation density  when its contribution
is larger  leading to a softer symmetry energy. 
The effects of the inclusion of $\delta$ mesons on the properties of compact stars, such as
mass, radius and strangeness content were
discussed in \cite{Menezes-04,Liu-05}. In \cite{Gaitanos-04} the $\delta$-meson effect on the density dependence of the symmetry
energy, on the  nucleon transport effects and on resonance and particle production around the threshold was studied. At subsaturation densities, the effect of the  $\delta$ meson
was investigated in the pasta formation at the inner crust of a compact star  \cite{Avancini-09},
the extension of the spinodal region \cite{Avancini-06} or the  spinodal region in the
presence of very strong magnetic fields as the ones that might occur in magnetars
\cite{Rabhi-arXiv09}.

The main goal of this work is to study the effect of $\delta$ mesons on the dynamical instabilities and phase transitions in nuclear matter within the framework of relativistic models investigating the effects of $\delta$ mesons on matter at finite temperatures and under the conditions of isospin asymmetry and density expected in the inner crust. It is important to try these models at different conditions in order to better understand the properties of the neutron star crust. Some works have treated clusterization of matter both in the non-relativistic  \cite{Chomaz-04,Shlomo-04} and relativistic \cite{ProvidenciaC-06,Lucilia-06} contexts. The relativistic Vlasov equation formalism applied in the last two papers will be used in the present work. In  \cite{Lucilia-06} warm nuclear  matter matter was studied within  NL3 \cite{Lalazissis-97}.
We use three different models -- NL3 \cite{Lalazissis-97}, NL$\rho\delta$ and NL$\rho$ \cite{Liu-02} -- to try to understand the role of $\delta$ mesons in that context. We will consider both neutral  neutron-proton-electron ($npe^-$) neutrino free matter in $\beta$-equilibrium at zero temperature and $npe^-$  matter with trapped neutrinos in $\beta$-equilibrium for a lepton fraction $Y_{Le}=0.4$. We will also stablish a comparison with the results discussed in \cite{Ducoin-08} for non-relativistic Skyrme models and the NL3 relativisitc model.

In Sec. \ref{II} we show the formalism we use and in Sec. \ref{IV} we present and discuss some of the results obtained and finally, in Sec. \ref{V}, some conclusions are taken.

\section{The Vlasov equation formalism} \label{II}

We use the relativistic non-linear Walecka model (NLWM) in the Mean-Field Approximation, within the Vlasov 
formalism to study nuclear collective modes of $npe^-$ matter at finite temperature \cite{Nielsen-91}. 

We consider a system of baryons, with mass $M$
interacting with and through an isoscalar-scalar field $\phi$ with mass
$m_s$, an isoscalar-vector field $V^{\mu}$ with mass
$m_v$, an isovector-scalar field $\boldsymbol \Delta$ with mass $m_{\delta}$ and an isovector-vector field $\mathbf b^{\mu}$ with mass $m_\rho$. We also include a system of electrons with mass $m_e$. Protons and electrons
interact through the electromagnetic field $A^{\mu}$.
The Lagrangian density reads:
$$
{\cal L}=\sum_{i=p,n} {\cal L}_i + {\cal L}_e + {\cal L}_{\sigma} + {\cal L}_{\omega} + {\cal L}_{\delta} + {\cal L}_{\rho} + {\cal L}_A
$$
where the nucleon Lagrangian reads
$$
{\cal L}_i=\bar \psi_i\left[\gamma_\mu i D^{\mu}-M^*\right]\psi_i \, ,
$$
with
$$
i D^{\mu}=i\partial^{\mu}-g_v V^{\mu}-
\frac{g_{\rho}}{2}  {\boldsymbol\tau} \cdot \mathbf{b}^\mu - e A^{\mu}
\frac{1+\tau_3}{2} \,,
$$
$$
M^*=M-g_s \phi -g_{\delta}{\boldsymbol\tau} \cdot \boldsymbol{\Delta} \, ,
$$
and the electron Lagrangian is given by
$$
{\cal L}_e=\bar \psi_e\left[\gamma_\mu\left(i\partial^{\mu} + e A^{\mu}\right)
-m_e\right]\psi_e .
$$
The isoscalar part is associated with the scalar sigma ($\sigma$) field $\phi$, and the vector omega ($\omega$) field $V_{\mu}$, whereas the isospin dependence comes from the isovector-scalar delta ($\delta$) field $\Delta^i$, and the isovector-vector rho ($\rho$) field $b_\mu^i$ (where $\mu$ stands for the four dimensional space-time indices  and $i$ the three-dimensional isospin direction index). The associated Lagrangians are:
\begin{eqnarray}
{\cal L}_\sigma&=&+\frac{1}{2}\left(\partial_{\mu}\phi\partial^{\mu}\phi
-m_s^2 \phi^2 - \frac{1}{3}\kappa \phi^3 -\frac{1}{12}\lambda\phi^4\right),\nonumber\\
{\cal L}_\omega&=&-\frac{1}{4}\Omega_{\mu\nu}\Omega^{\mu\nu}+\frac{1}{2}
m_v^2 V_{\mu}V^{\mu}, \nonumber \\
{\cal L}_\delta&=&+\frac{1}{2}\partial_{\mu}\boldsymbol \Delta\partial^{\mu}\boldsymbol \Delta
-\frac{1}{2}m_{\delta}^2 \boldsymbol \Delta^2, \nonumber\\
{\cal L}_\rho&=&-\frac{1}{4}\mathbf B_{\mu\nu}\cdot\mathbf B^{\mu\nu}+\frac{1}{2}
m_\rho^2 \mathbf b_{\mu}\cdot \mathbf b^{\mu}, \nonumber\\
{\cal L}_A&=&-\frac{1}{4}F_{\mu\nu}F^{\mu\nu},
\end{eqnarray}
where
$\Omega_{\mu\nu}=\partial_{\mu}V_{\nu}-\partial_{\nu}V_{\mu} ,
\quad \mathbf B_{\mu\nu}=\partial_{\mu}\mathbf b_{\nu}-\partial_{\nu} \mathbf b_{\mu}
- g_\rho (\mathbf b_\mu \times \mathbf b_\nu)$ and $F_{\mu\nu}=\partial_{\mu}A_{\nu}-\partial_{\nu}A_{\mu}$.

The model comprises the following parameters:
four coupling constants $g_s$, $g_v$, $g_{\delta}$ and $g_{\rho}$ of the mesons to
the nucleons, the bare nucleon mass $M$, the electron mass $m_e$, the masses of
the mesons, the electromagnetic coupling constant
$e=\sqrt{4 \pi/137}$ and
the self-interacting coupling constants $\kappa$ and $\lambda$. In this Lagrangian density, $\boldsymbol \tau$ is the isospin operator.

We have used the  set of constants I and II, identified here as NL$\rho$ and NL$\rho\delta$, respectively, taken from \cite{Liu-02} and also the NL3 parametrization \cite{Lalazissis-97}. In the first two cases, the saturation density that we refer as $\rho_0$ is 0.160 fm$^{-3}$. For NL3, $\rho_0=0.148$ fm$^{-3}$. 

Table \ref{7} shows some nuclear matter properties at the saturation density for these models: the binding energy per nucleon, the incompressibility coefficient, the symmetry energy, the symmetry energy slope, the symmetry energy curvature and $K_{\tau}=K_{sym}-6L-\frac{Q_0}{K}L$.

All these properties, except $K_{sym}$ and $K_{\tau}$, present lower values for NL$\rho$ and NL$\rho\delta$, as compared to NL3. The two parametrizations, NL$\rho$ and NL$\rho\delta$, have $L$ values that are within the expected range estimated from recent experimental constraints: $L=88\pm25$ MeV \cite{Xu-08,Xu-09}. However, the values of $K_{\tau}$ fall out of the $K_{\tau}=-550\pm100$ MeV \cite{Li-07,Garg-07} range.

\begin{table}
  \begin{tabular}{|c|c|c|c|c|c|c|c|}
    \hline
 Model&$\rho_{0}$&$E/A$&$K$&$E_{sym}$&$L$&$K_{sym}$&$K_{\tau}$   \\
    \hline
NL3&0.148&-16.240&269.936&37.344&118.320&100.525&-696.129\\
NL$\rho$&0.160&-16.051&239.884&30.335&84.510&3.328&-340.293\\
NL$\rho\delta$&0.160&-16.051&239.884&30.711&102.664&127.246&-290.187\\
    \hline
  \end{tabular}
\caption{ Nuclear matter properties at saturation density, $\rho_0$. All quantities are in MeV, except for $\rho_0$, given in fm$^{-3}$. } \label{7}  
\end{table}

We denote by
$$f({\bf r},{\bf p},t)_\pm=\mbox{diag}(f_{p \pm},\,f_{n \pm},\,f_{e \pm})$$
the  distribution functions of particles (+) at position $\mathbf r$,
instant $t$ and momentum $\mathbf{p}$ and of antiparticles (-) at position $\mathbf r$,
instant $t$ and momentum $-\mathbf{p}$,
and by
\begin{equation}
h_{\pm}=\mbox{diag}(h_{p\pm},h_{n\pm},h_{e\pm})
\end{equation}
the corresponding one-body hamiltonian, where
\begin{equation}
h_{i\pm}=\pm \sqrt{({\bf p}-{\boldsymbol{\cal V}}_i)^2+M_i^{*2}}+ {\cal V}_{0i}.
\label{aga}
\end{equation}
For protons and neutrons, $i=p,n$, we have
\begin{eqnarray*}
{\cal V}_{0i}&=& g_v V_0  + \frac{g_\rho}{2}\tau_i b_0+ e A_0 \frac{1+\tau_{i}}{2} ,\\
{\boldsymbol{{\cal V}}}_{i}&=& g_v {\mathbf V} +
\frac{g_\rho}{2}\, \tau_i {\mathbf b}+ e\,  {\mathbf A} \frac{1+\tau_{i}}{2}, \\
 M_i^*&=&M-g_s\phi-g_{\delta}\tau_i\Delta_{3}
\end{eqnarray*}
with $\tau_i=1$ (protons) or -1 (neutrons). For electrons, $i=e$, we have
$$
{\cal V}_{0e}=-eA_0,\qquad {\boldsymbol {\cal V}}_e=-e {\bf A},\qquad M_e^*=m_e.
$$

The time evolution of the distribution functions is described by the
Vlasov equation
\begin{equation}
\frac{\partial f_{i \pm}}{\partial t} +\{f_{i \pm},h_{i \pm}\}=0, \qquad 
\; i=p,\,n,\, e,
\label{vlasov1}
\end{equation}
where $\{,\}$ denote the Poisson brackets.

From Hamilton's equations we derive the equations describing the time
evolution of the fields $\phi$,  $V^\mu$, $A^\mu$, the third component of
the $\rho$-field, $b_3^\mu=(b_0,\mathbf{b})$, and the third isospin component, $\Delta_3$, of the $\boldsymbol \Delta$ field, which are given in Appendix \ref{eq_campos}.

The state which minimizes the energy of asymmetric nuclear matter
is characterized by the distribution functions
\begin{equation*}
f_{0 i \pm}= \frac{1}{1+e^{(\epsilon_{0 i} \mp \nu_i)/T}}, \quad i=p,n,
\end{equation*}
with
\begin{equation*}
\epsilon_{0 i}=\sqrt{p^2+{M_i^*}^2},  \quad \nu_i=\mu_i - g_v V_0  - \frac{g_\rho}{2}\, \tau_i b_0 - e\, A_0 
\frac{1+\tau_i}{2} 
\end{equation*}
and
\begin{equation*}
f_{0 e \pm}= \frac{1}{1+e^{(\epsilon_{0 e} \mp \mu_e)/T}} \,\, ,
\end{equation*}
with
\begin{equation*}
\epsilon_{0e}=\sqrt{p^2+m_e^2}
\end{equation*}
and by the  constant mesonic fields, which obey the following equations:
$m_s^2\phi_0^{(0)} + \frac{\kappa}{2} \phi_0^{(0)\,2} + \frac{\lambda}{6} \phi_0^{(0)\,3}=g_s\rho_s^{(0)},\,\,\,$
$m_v^2\,V_0^{(0)}=g_v j_0^ {(0)},\,\,\,$
$m_{\rho}^2\,b_0^{(0)}=\frac{g_\rho}{2} j_{3,0}^{(0)},\,\,\,$
$m_\delta^2\Delta^{(0)}_3=g_\delta \rho_{3,s}^{(0)},\,\,\,$
$V^{(0)}_i=b_i^{(0)}= A_0^{(0)}= A_i^{(0)}=0.\,\,$

Collective modes in the present approach correspond to small oscillations
around the equilibrium state. These small deviations are described by the 
linearized equations of motion and, therefore, collective modes are given as 
solutions of those equations. To construct them, let us define: 
\begin{eqnarray*}
f_{i \pm}&=&f_{0 i \pm}^{(0)} + \delta f_{i \pm}, \\ \nonumber
\phi_0\,&=&\,\phi_0^{(0)} + \delta\phi, \\ \nonumber
\Delta_3\,&=&\,\Delta^{(0)}_3 + \delta \Delta_3, \\ \nonumber
V_0\,&=&\, V_0^{(0)} + \delta V_0 , \quad V_i\,=\,\delta V_i, \\ \nonumber
b_0\,&=&\, b_0^{(0)} + \delta b_0, \quad b_i\,=\,\delta b_i, \\ \nonumber
A_0\,&=&\, \delta A_0, \quad A_i\,=\,\delta A_i.
\end{eqnarray*}

As in \cite{Nielsen-91,Avancini-05,Avancini-04}, we
express the fluctuations of the distribution functions in terms of the  generating 
functions:
$$S_{\pm}({\mathbf r},{\mathbf p},t)=
\mbox{diag}\left(S_{p \pm},\,S_{n \pm},\,S_{e \pm}\right),$$
such that 
$$\delta f_\pm=\{S_\pm,f_{0,\pm}\}=\{S_\pm,p^2\}\frac{df_{0 \pm}}{dp^2}.$$
In terms of the generating functions, the linearized Vlasov equations
for $\delta f_{i \pm}$,
$$\frac{d\delta f_{i \pm}}{d t}+ \{\delta f_{i \pm}, h_{0 i \pm}\}
 +\{f_{0 i \pm},\delta h_{i \pm} \}=0$$
 are equivalent to the following time-evolution equations:

\begin{eqnarray*}
\label{eq:deltaf}
\frac{\partial S_{i \pm}}{\partial t} + \{S_{i \pm},h_{0i \pm}\} &=& \delta h_{i \pm} = \mp \frac{g_s\, M_i^*}{\epsilon_{0 i}}\delta\phi 
\mp \frac{g_\delta\, \tau_i\, M_i^*}{\epsilon_{0 i}}\delta \Delta_3 \\ \nonumber
&\mp& \frac{{\bf p} \cdot \delta \boldsymbol{\cal V}_i}{\epsilon_{0 i}}  + \delta{\cal V}_{0i}, \quad i=p,n
\end{eqnarray*}
\begin{equation*}
  \label{eq:deltafe}
\frac{\partial S_{e \pm}}{\partial t} + \{S_{e \pm},h_{0e \pm}\} = \delta h_{e \pm} = -e\left[ \delta{A}_{0} \mp \frac{{\bf p} \cdot \delta{\mathbf A}}{\epsilon_{0e}}\right], 
\end{equation*}
where
\begin{eqnarray*}
\delta{\cal V}_{0i}&=&g_v \delta V_0 + \tau_i \frac{g_{\rho}}{2}\, 
\delta b_0 + e\, \frac{1+\tau_{i}}{2}\,\delta A_0 ,\\ \nonumber
\delta \boldsymbol{\cal V}_i&=& g_v \delta {\mathbf V} + \tau_i
 \frac{g_{\rho}}{2} \,\delta {\mathbf b}+ e\, \frac{1+\tau_{i}}{2}\,
\delta {\mathbf A},
\end{eqnarray*}
with $ h_{0 i \pm}\,= \pm \epsilon_{0 i} +{\cal V}^{(0)}_{0 i}$ and $h_{0 e \pm}=\pm \epsilon_{0 e}$.

Of particular interest on account of their physical relevance are
the longitudinal modes, with momentum ${\bf k}$ and frequency $\omega$,
described by the ansatz
$$
\left(\begin{array}{c}
S_{j \pm}({\bf r},{\bf p},t)  \\
\delta\phi  \\ 
\delta \Delta_3 \\
\delta \xi_0 \\ \delta \xi_i
\end{array}  \right) =
\left(\begin{array}{c}
{\cal S}_{\omega\pm}^j (p,{\rm cos}\theta) \\
\delta\phi_\omega \\
\delta \Delta_{3\,\omega} \\
\delta \xi_\omega^0\\ \delta \xi_\omega^i 
\end{array} \right) {\rm e}^{i(\omega t - {\bf k}\cdot
{\bf r})} \;  ,
$$
where $j=p,\, n,\, e$, $\xi=V,\, b,\, A$ represents the vector-meson fields and $\theta$ is the angle between ${\bf p}$ and ${\bf k}$. For these modes,
we get $\delta V_\omega^x = \delta V_\omega^y =0\,$, $\delta b_\omega^x =
\delta b_\omega^y = 0\,$ and $\delta A_\omega^x = \delta A_\omega^y
=0\,$. \\ 
Calling $\delta \Delta_{3\,\omega}=\delta \Delta_\omega$, $\delta V_\omega^z = \delta V_\omega$, $\delta b_\omega^z = \delta 
b_\omega$ and $\delta A_\omega^z = \delta A_\omega$, we will  have
$ \delta {\cal V}_{i,z}= \delta {\cal V}_\omega^i{\rm e}^{i(\omega t - {\bf k}\cdot
{\bf r})}$ and \\ $\delta {\cal V}_{0i}= \delta {\cal V}_\omega^{0i}{\rm e}^{i(\omega t - {\bf k}\cdot
{\bf r})}$. In the Appendix we present the equations for the fields (eqs. (\ref{eqmphi})-(\ref{eqma})), the equations of motion for the fluctuations (eqs. (\ref{Se})-(\ref{A})), the solutions for the eigenmodes (\ref{matriz}), as well as defining the coefficients $a_{ij}$. There we also show the amplitudes $\rho^S_{\omega i}$ (eq.(\ref{rhosi})) and $\rho_{\omega i}$ (eq.(\ref{rhoi})) obtained from the dispersion relation (eq. \ref{det}).

\section{Numerical results and discussions} \label{IV}

From the dispersion relation (see Appendix \ref{rel_disp}, eq.(\ref{det})) we have obtained the dynamical spinodal surfaces,
characterized by a zero frequency, for different temperatures and momentum
transfer, the ratios of the proton to neutron  transitions
densities (see eq.(\ref{amp1})) which allow us to discuss the distillation effect. We also show the unstable modes with the largest energy modulus, which defines the mode that drives the
system to a non-homogeneous phase and, therefore, gives an estimation of the
size of the clusters formed in the phase transition. In the sequel, we will discuss the effect of
$\delta$ mesons in these three physical quantities.

\begin{figure*}
 \begin{tabular}{ll}
   \begin{tabular}{c}
   \includegraphics[height=5.cm]{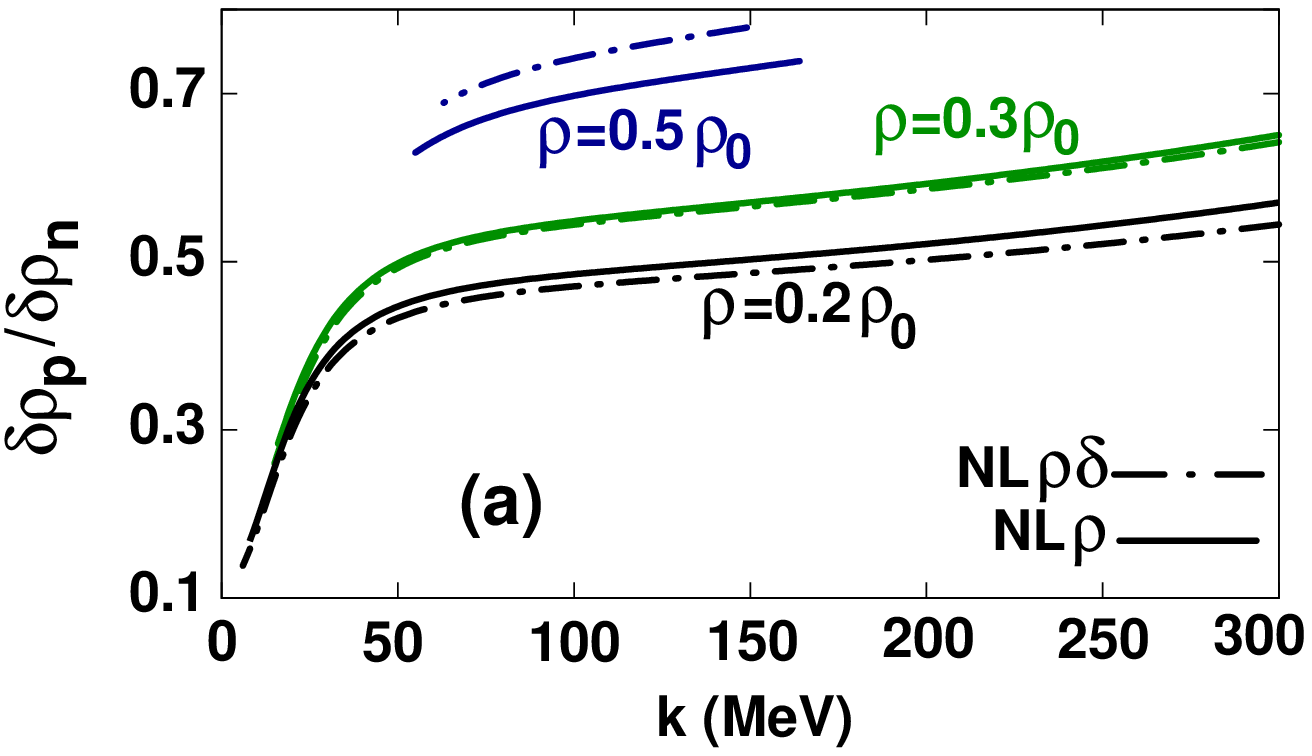}
   \end{tabular}
   \begin{tabular}{c}
   \includegraphics[height=5.cm]{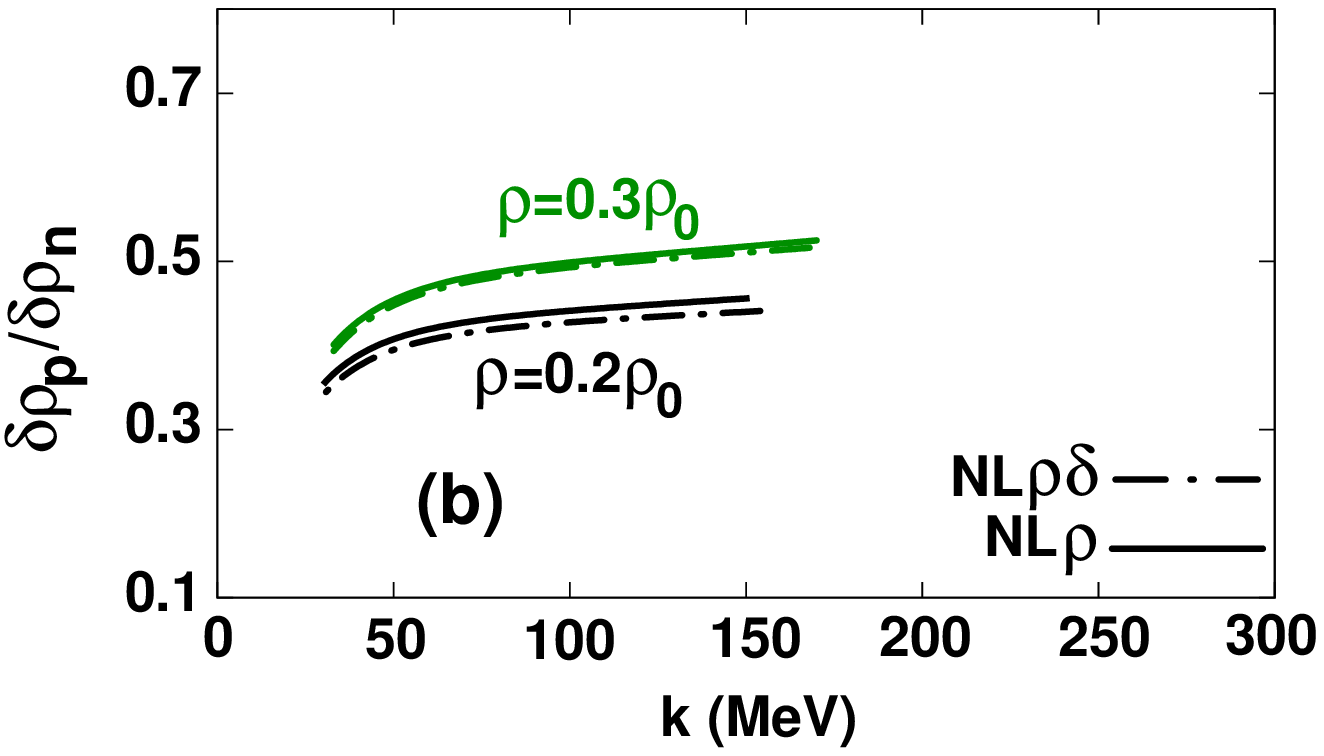}
   \end{tabular}
 \end{tabular}
\caption{(Color online) The ratio of the proton over the neutron density fluctuations plotted for $y_p=0.2$ and $\rho=0.2 \rho_0$ (black), $\rho=0.3 \rho_0$ (green) and $\rho=0.5 \rho_0$ (blue) at $T=5$ (a) and $T=10$ MeV (b) as a function of the transferred momentum for NL$\rho\delta$ (dot-dashed) and NL$\rho$ (continuous line).}%
\label{fig1}
\end{figure*}

Figure \ref{fig1} shows the ratio of the proton to neutron density fluctuations
as a function of the transferred momentum at $T=5$ and 10 MeV. At $T=5$ MeV
and for the largest density considered, i.e., $\rho=0.5\, \rho_0$, the $\delta$
meson gives rise to a larger distillation effect. However, for $\rho=0.3
\rho_0$ the effect of $\delta$ is negligible and for $\rho=0.2 \rho_0$ it
reduces the distillation effect. This trend is not modified with temperature,
except that the instability for the larger densities may disappear.

\begin{figure*}
 \begin{tabular}{ll}
   \begin{tabular}{c}
   \includegraphics[height=5.cm]{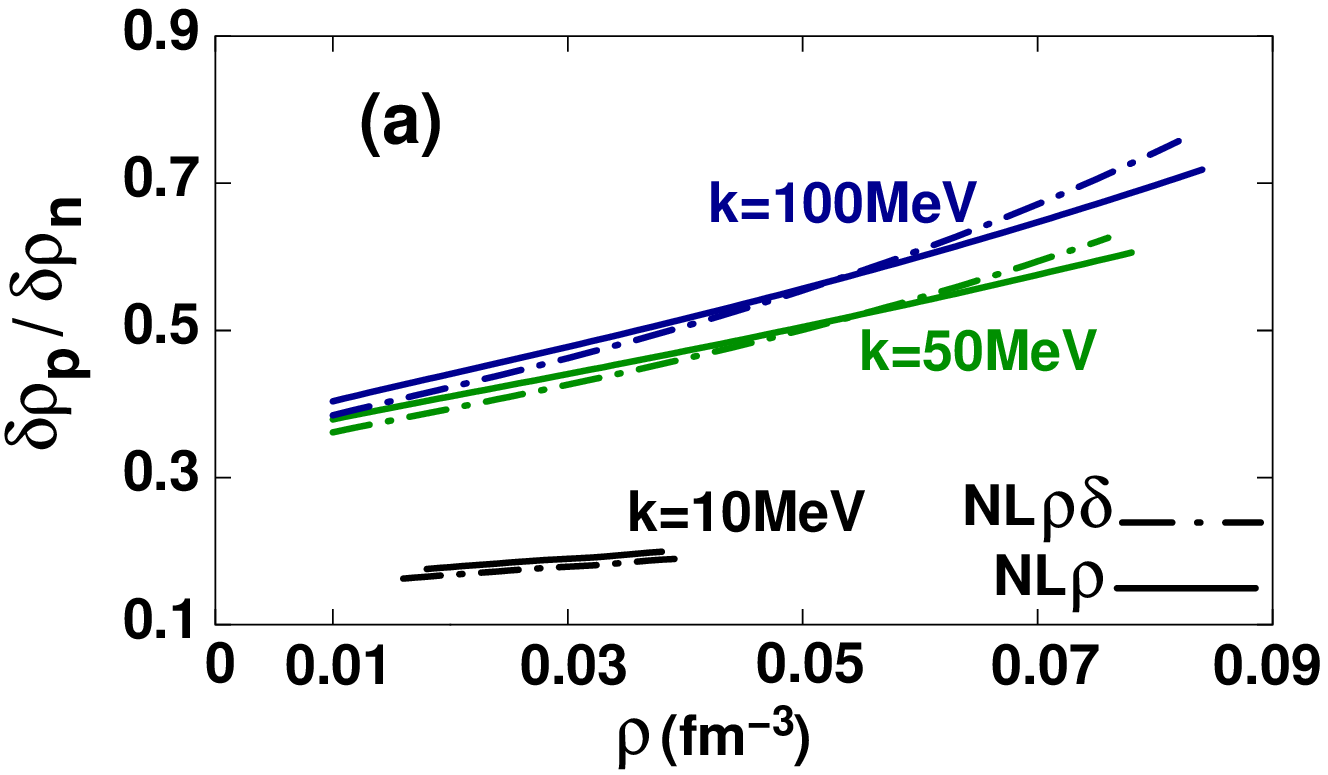}
   \end{tabular}
   \begin{tabular}{c}
   \includegraphics[height=5.cm]{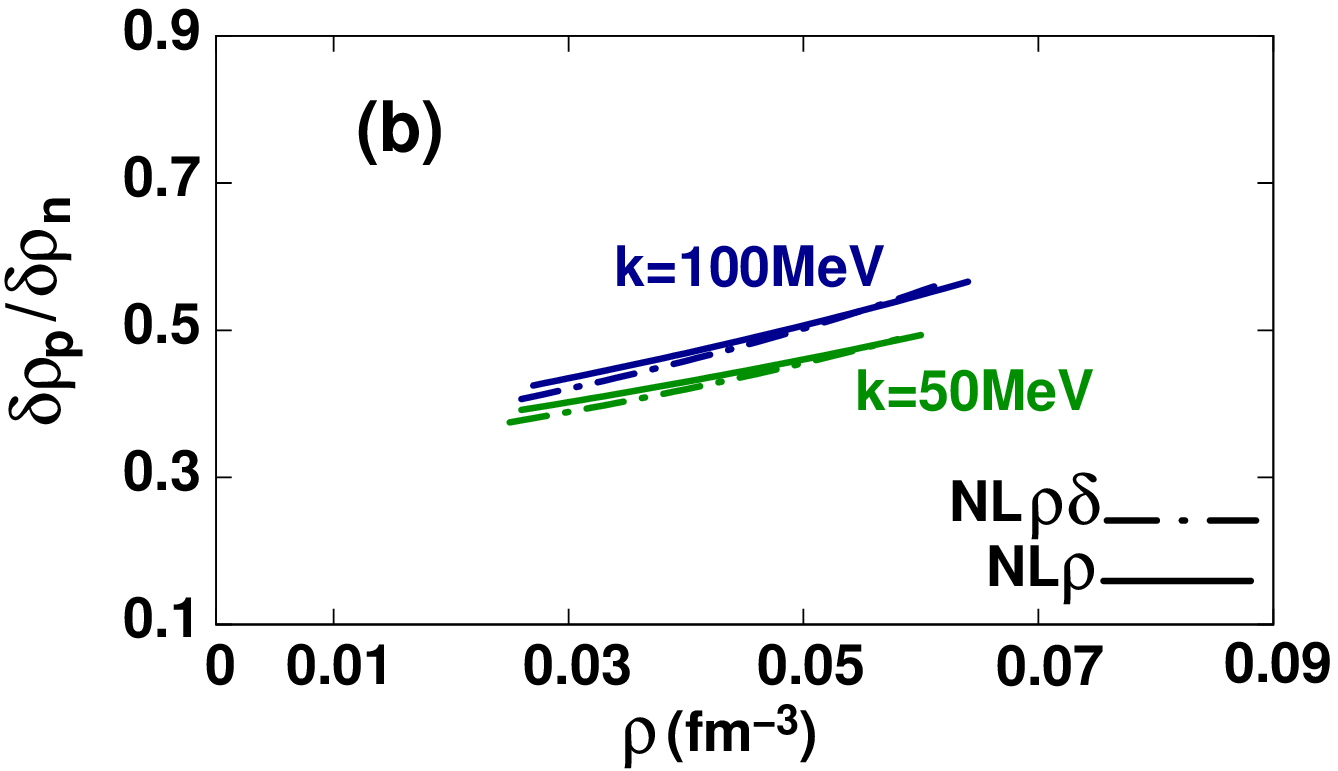}
   \end{tabular}
 \end{tabular}
\caption{(Color online) The ratio of the proton over the neutron density fluctuations plotted for $y_p=0.2$ and $k=10$ MeV (black), $k=50$ MeV (green) and $k=100$ MeV (blue) at $T=5$ (a) and $T=10$ MeV (b) as a function of the density for NL$\rho\delta$ (dot-dashed) and NL$\rho$ (continuous line).}%
\label{fig2}
\end{figure*}

This effect is better seen from Fig. \ref{fig2} where the ratio of the proton to
neutron density fluctuations are also plotted  as a function of the density
for several values of the transferred momentum. As  in Fig. \ref{fig1}, we  see
that for a density above (below) $\rho \sim 0.06$ fm$^{-3}$, the $\delta$ meson
increases  (decreases) the
distillation effect. This will affect the constitution of the crust: in the region closer to
the inner edge the clusters are more proton rich, whereas the gas is more neutron-rich than the prevision without considering  $\delta$ mesons. For lower densities, the opposite happens.

\begin{figure*}
 \begin{tabular}{ll}
   \begin{tabular}{cc}
   \includegraphics[height=5.5cm]{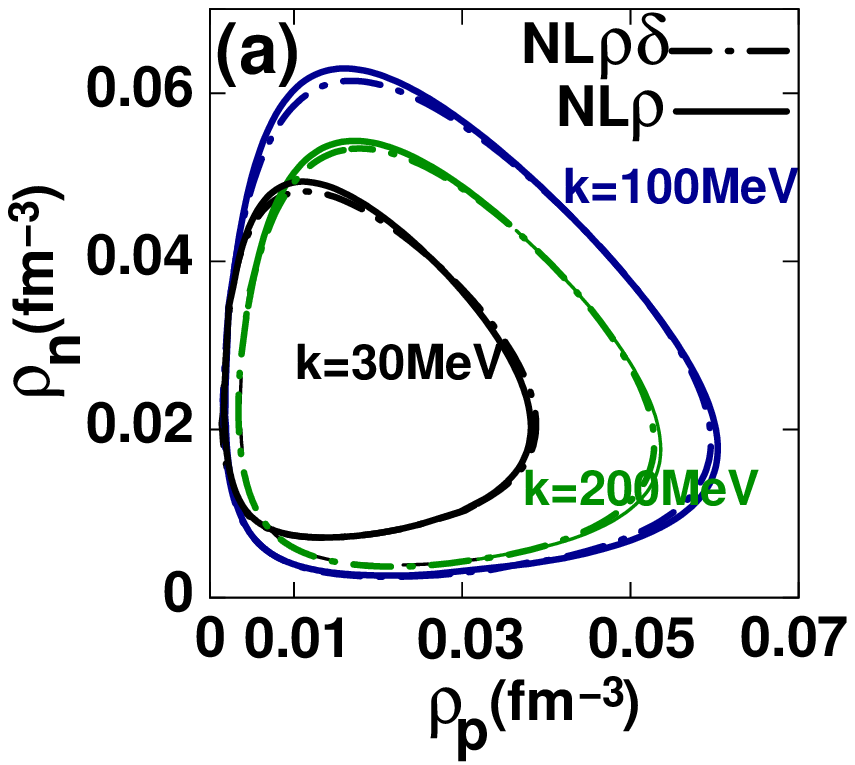}\\
   \includegraphics[height=5.5cm]{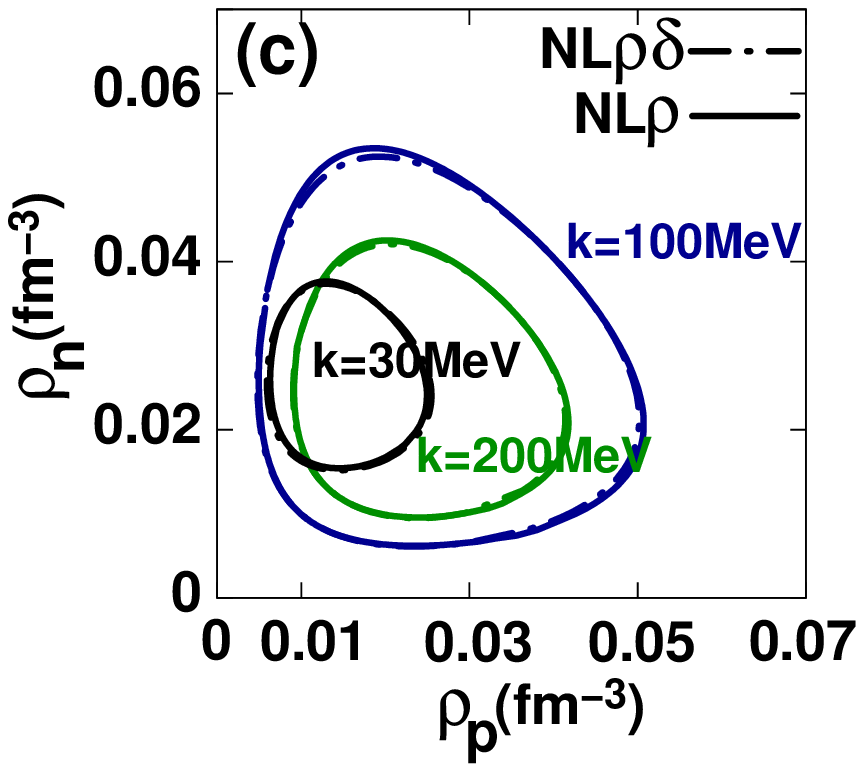}
   \end{tabular}
   \begin{tabular}{cc}
   \includegraphics[height=5.5cm]{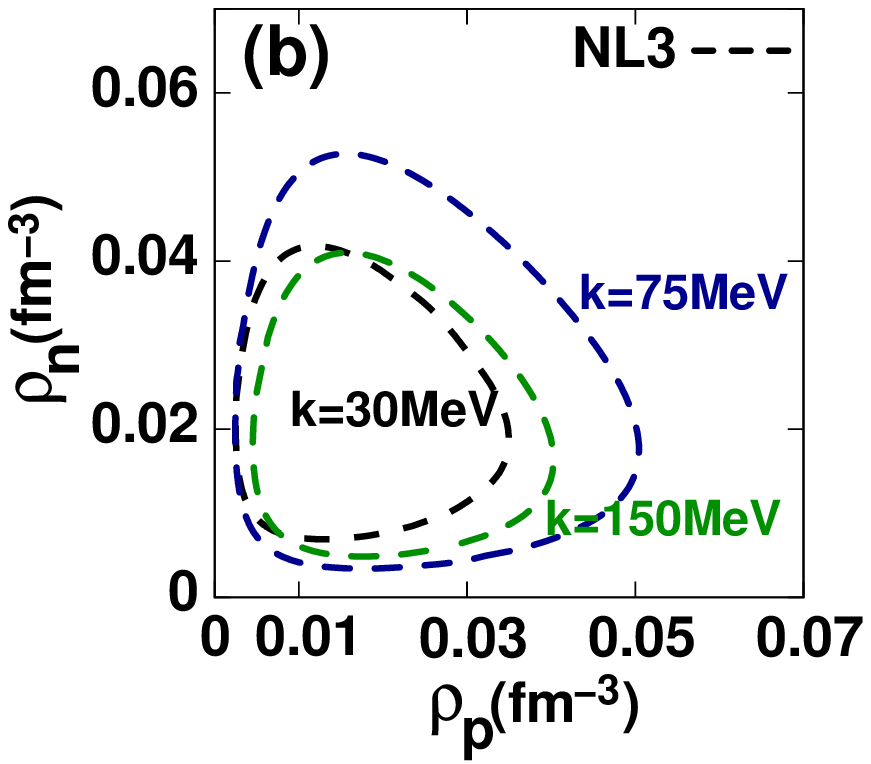}\\
   \includegraphics[height=5.5cm]{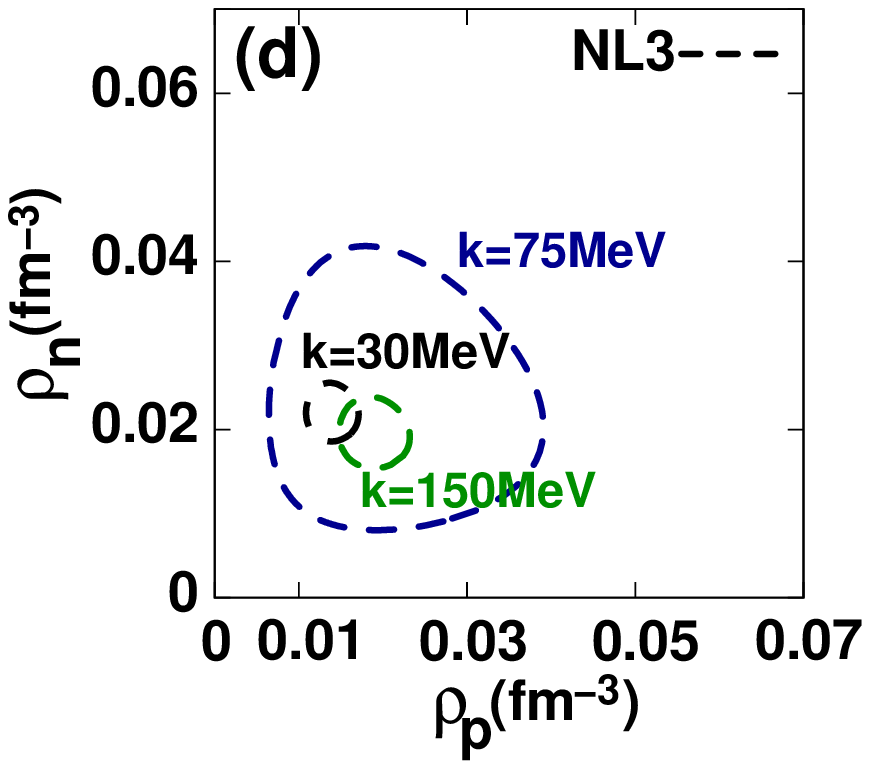}
   \end{tabular}
 \end{tabular}
\caption{(Color online) Spinodal curves for $k=30$ MeV (black), $k=75$ and $100$ MeV (blue), $k=150$ and $200$ MeV (green) at $T = 7$ ((a) and (b)) and $T=10$ MeV ((c) and (d)) for NL$\rho\delta$ (dot-dashed), NL$\rho$ (continuous) and NL3 (dashed line).}%
\label{fig3}
\end{figure*}

Figure \ref{fig3} shows the subsaturation  instability regions at $T=7$ and 10
MeV for NL$\rho\delta$, NL$\rho$ and NL3. It can be seen that these regions
become smaller with the increase of temperature.  The presence of the $\delta$
meson also reduces slightly the spinodal region for very isospin asymmetric
matter. For NL$\rho\delta$ and NL$\rho$, the envelope of the spinodal region is
characterized by $k\sim 100$ MeV.  On the other hand, NL3 has a smaller
unstable region as compared with the other two parametrizations and the spinodal
envelope is obtained for a smaller $k$, $k\sim75$ MeV.
The different $k$ dependencies originate in the finite-range
part of the nuclear force as discussed in \cite{Ducoin-08}.

\begin{figure*}
 \begin{tabular}{ll}
   \begin{tabular}{cc}
   \includegraphics[height=5.5cm]{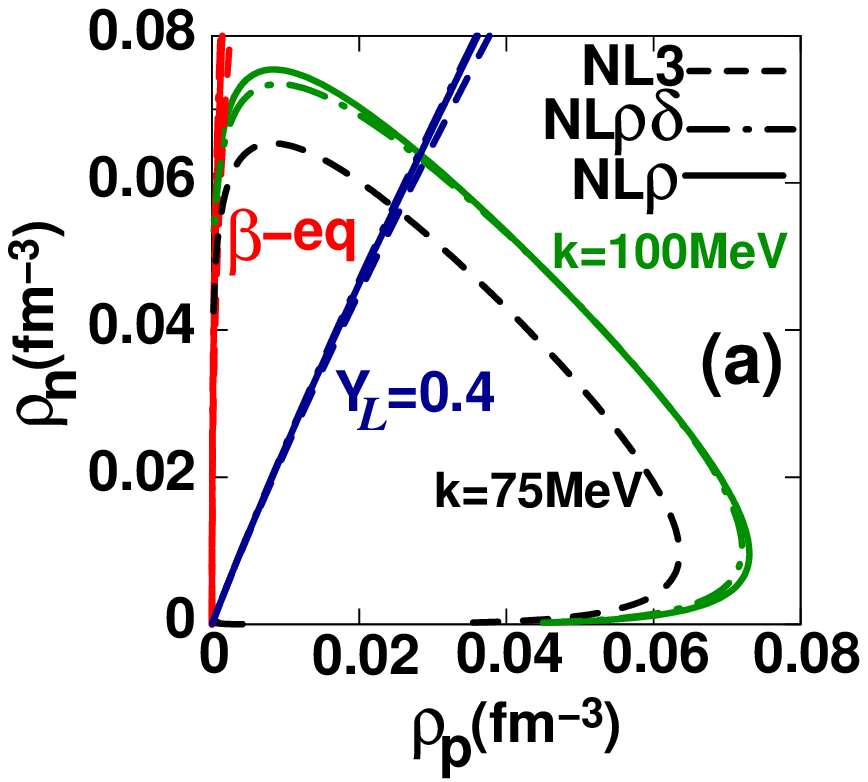}
   \includegraphics[height=5.5cm]{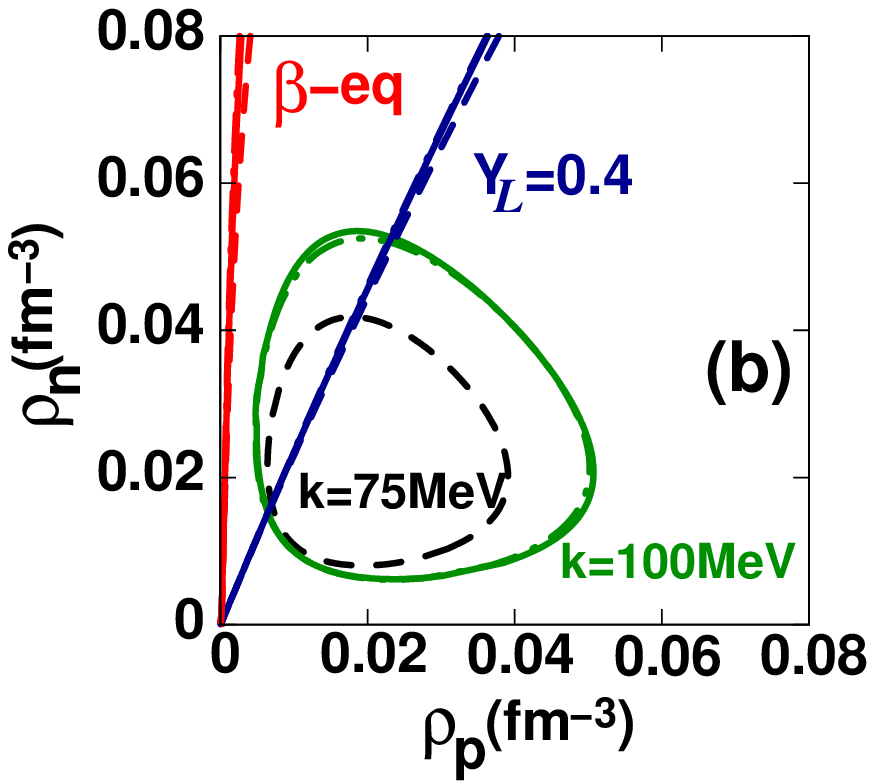}
   \end{tabular}
   \begin{tabular}{c}
   \includegraphics[height=5.5cm]{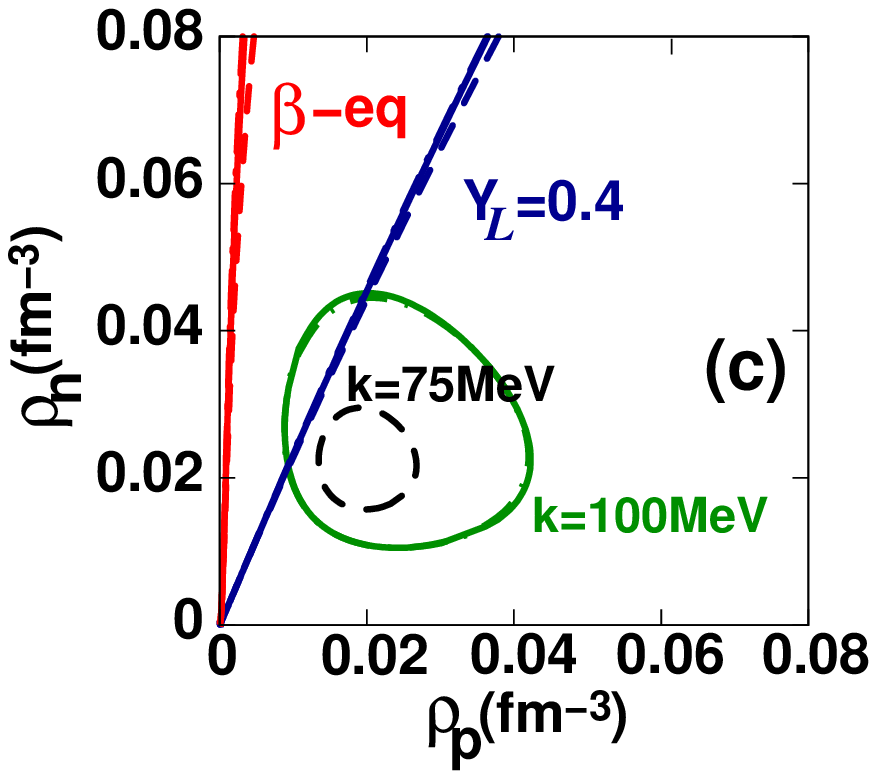}
   \end{tabular}
 \end{tabular}
\caption{(Color online) Spinodals for $k=75$ (black) and $k=100$ MeV (green) for NL3 (dashed), NL$\rho\delta$ (dot-dashed) and NL$\rho$ (continuous line) at $T=0$ (a), $T=10$ (b) and $T=12$ MeV (c). The lines that cross the spinodals are the EoS in $\beta$-equilibrium for $npe^-$ (neutrino-free) and with trapped neutrinos for a lepton fraction $Y_L=0.4$ (red and blue, respectively), as indicated.}%
\label{fig4}
\end{figure*}

Figure \ref{fig4} shows the approximate envelope of  the spinodal regions for different temperatures and
the three models considered. The lines that cross the neutron-rich part of the spinodal are the EoS
for $\beta$-equilibrium matter both without neutrinos and considering neutrino-trapping with a lepton fraction $Y_L=0.4$, as indicated in the figure. The non-homogeneous phase describing the crust corresponds to the EoS inside the spinodal. We notice that NL$\rho$ and NL$\rho\delta$ are almost coincident and only differ at large isospin asymmetries. We can see that the EoS for $\beta$-equilibrium without neutrinos cross the spinodal region at $T=0$ MeV, and small temperatures. At $T>3$ MeV, the non-homogeneous phase disappears in neutrino-free stellar matter. It means that after the neutrino outflow, a non homogenous phase appears at the crust of the star when the star's surface has cooled down to a temperature below $\sim 3$ MeV. For matter with trapped neutrinos, the EoS cross the spinodal for $T<11.6$ MeV, considering the NL3 parametrization and $T<13.2$ MeV, for the NL$\rho\delta$ and NL$\rho$ parametrizations. The non-homogeneous phase in the crust of the proto-neutron star will,
therefore,  exist if the temperature of the crust is not larger than $\sim 12$
MeV being this critically model dependent.

\begin{figure}
   \begin{tabular}{c}
   \includegraphics[height=5cm]{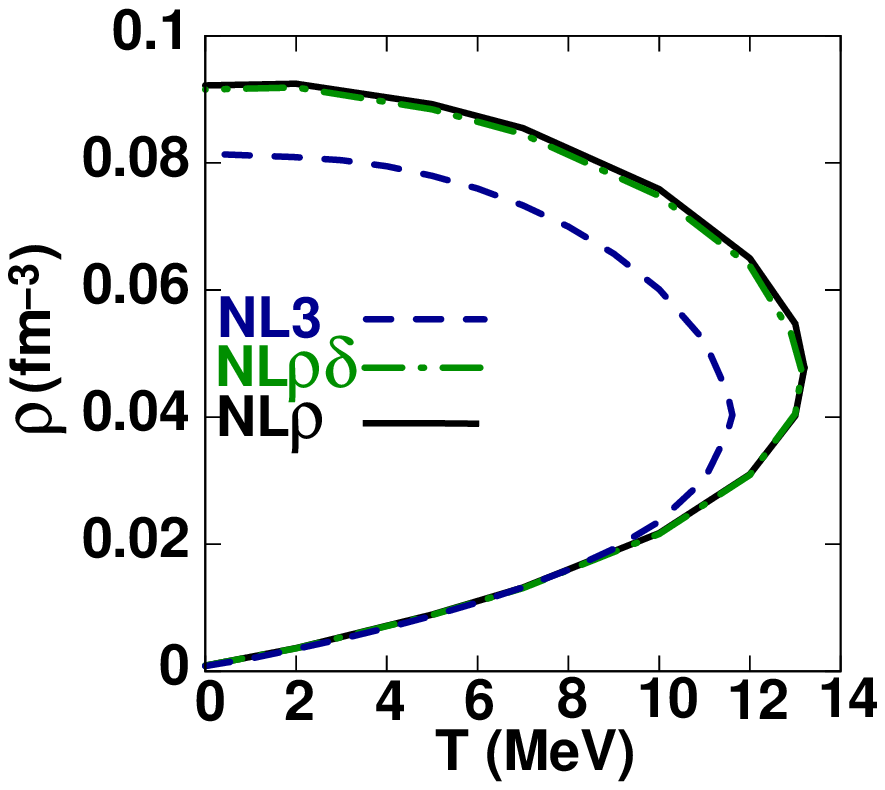}
   \end{tabular}
\caption{(Color online) The crossing densities at the crust-core transition for $\beta$-equilibrium stellar matter with trapped neutrinos ($Y_L=0.4$) for NL3 (blue, dashed), NL$\rho\delta$ (green, dot-dashed) and NL$\rho$ (black, continuous line) as a function of temperature.}%
\label{fig5}
\end{figure}

In Fig. \ref{fig5} the transition densities at the crust-core transition for $\beta$-equilibrium stellar matter with trapped neutrinos ($Y_L=0.4$) are plotted as a function of temperature. Those values, as well as the corresponding pressures for different temperatures are also shown in Table \ref{tab2}. The crust-core transition densities decrease with temperature. Also, the pressure shows a similar behaviour, except for NL3 where it slightly increases. Pressures are slightly larger in the presence of $\delta$ mesons.

In \cite{Ducoin-08}, the authors compare predictions from both nonrelativistic (Skyrme forces) and Relativistic Mean-Field (RMF) and density-dependent models (DDM). For Skyrme models, the crossing density for matter with trapped  neutrinos at T=0 occurs $\sim 0.088$ fm$^{-3}$, except for two special cases, just slightly larger than
most of the relativistic models studied and slightly below NL$\rho$ and NL$\rho\delta$. The critical temperature above which the instabilities disappear is $\sim$ 12 MeV in agreement with the temperature obtained for NL$\rho$ and NL$\rho\delta$.

\begin{table}[h]
  \begin{tabular}{|c|c c c|c c c|}
    \hline
    T[MeV]  & & $\rho_{cross}$[fm$^{-3}$] & &  &P[MeV.fm$^{-3}$] &   \\
    \hline
      & NL3 & NL$\rho\delta$ & NL$\rho$ & NL3 & NL$\rho\delta$ & NL$\rho$\\
    \hline
      0 & 0.081 & 0.091 & 0.092 & 0.995 & 1.254 & 1.242\\
    \hline
      2 & 0.079 & 0.092 & 0.092  & 0.969 & 1.256 & 1.245\\
    \hline
      5 & 0.078 & 0.088 & 0.089  & 1.087 & 1.250 & 1.242\\
    \hline
      7 & 0.073 & 0.084 & 0.085  & 1.105 & 1.249 & 1.244\\
    \hline
      10 & 0.060 & 0.075 & 0.076  & 0.971 & 1.244 & 1.243\\
    \hline
      12  & -- & 0.064 & 0.065 & -- & 1.122 & 1.124 \\
    \hline
      13  & -- & 0.053 & 0.055 & -- & 0.980 & 0.982 \\
    \hline
  \end{tabular}
\caption{The crossing densities and pressures at the crust-core transition for
  $\beta$-equilibrium stellar matter with trapped neutrinos ($Y_L=0.4$) for
  several temperature and the three models: NL3, NL$\rho\delta$ and NL$\rho$. } \label{tab2}
  \end{table}

\begin{figure}
 \begin{tabular}{l}
   \begin{tabular}{c}
   \includegraphics[height=5cm]{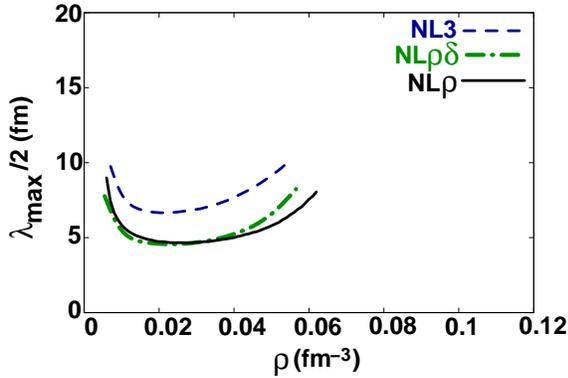}
   \end{tabular}
  \end{tabular}
\caption{(Color online) Size of the clusters in $\beta$-equilibrium in neutrino-free matter at $T = 0$ MeV for NL3 (blue, dashed), NL$\rho\delta$ (green, dot-dashed) and NL$\rho$ (black, continuous line).}%
\label{fig6}
\end{figure}
\begin{figure*}[!htb]
 \begin{tabular}{lll}
   \begin{tabular}{c}
   \includegraphics[height=4.cm]{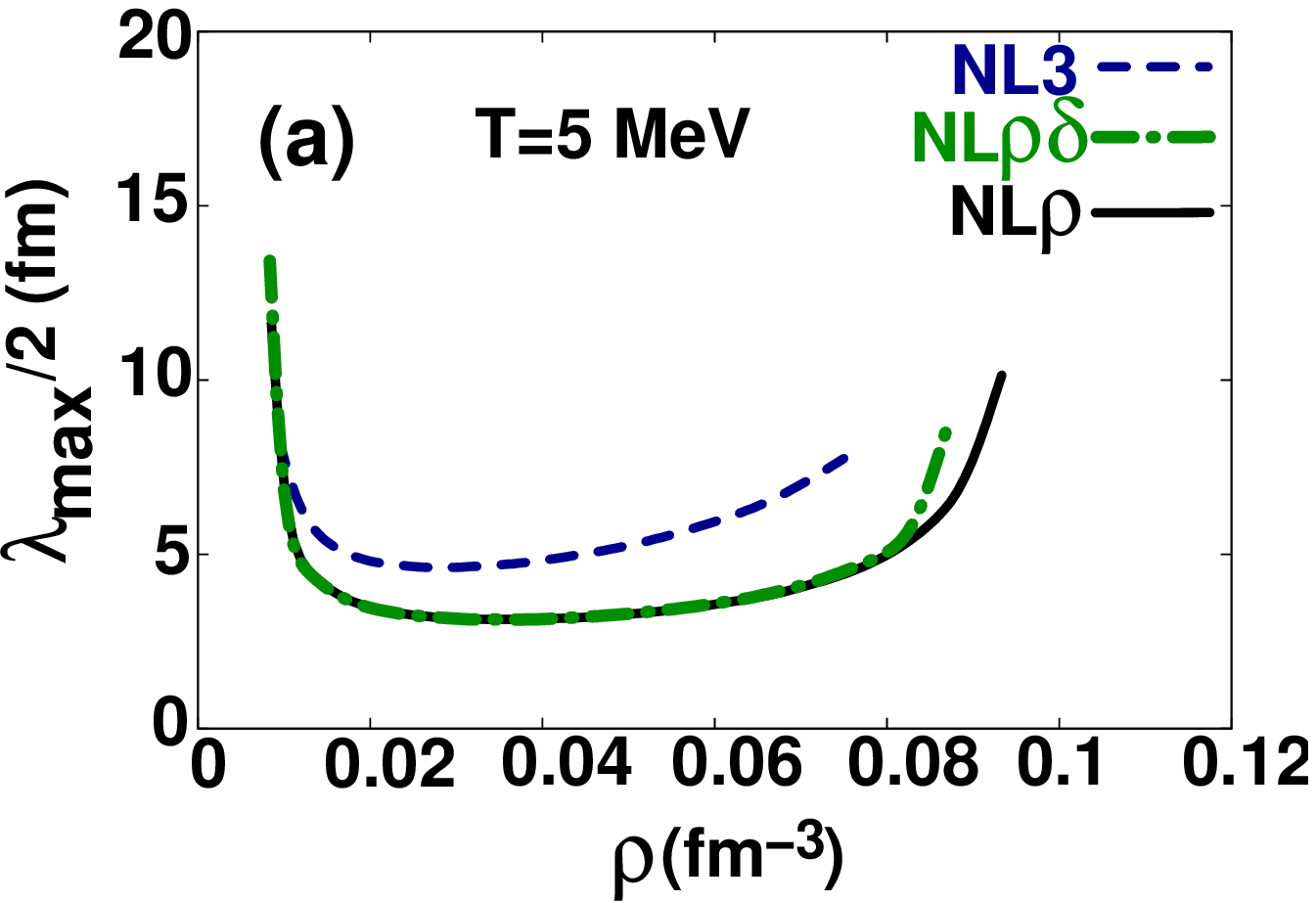}
   \end{tabular}
   \begin{tabular}{c}
   \includegraphics[height=4.cm]{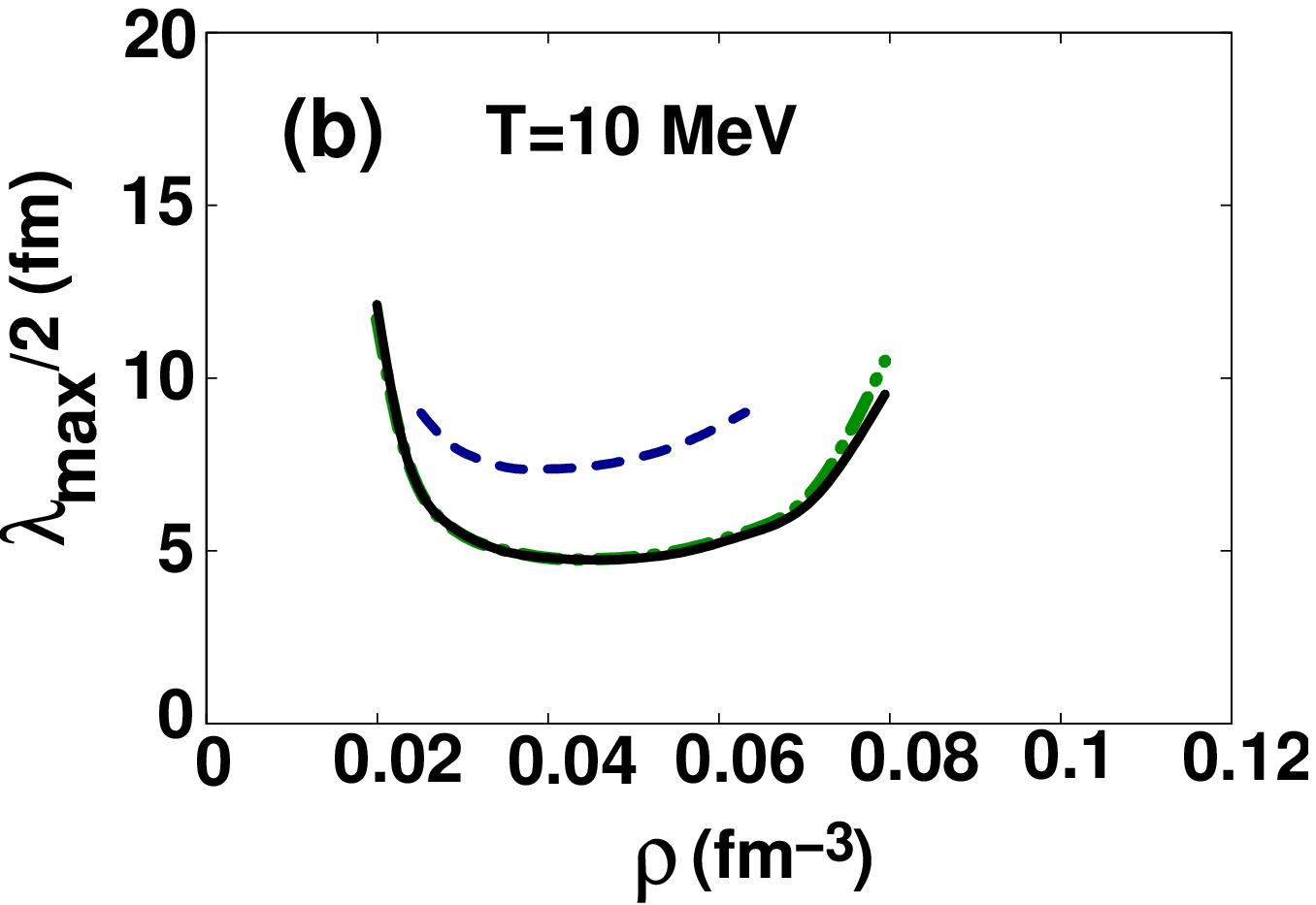}
   \end{tabular}
   \begin{tabular}{c}
   \includegraphics[height=4.cm]{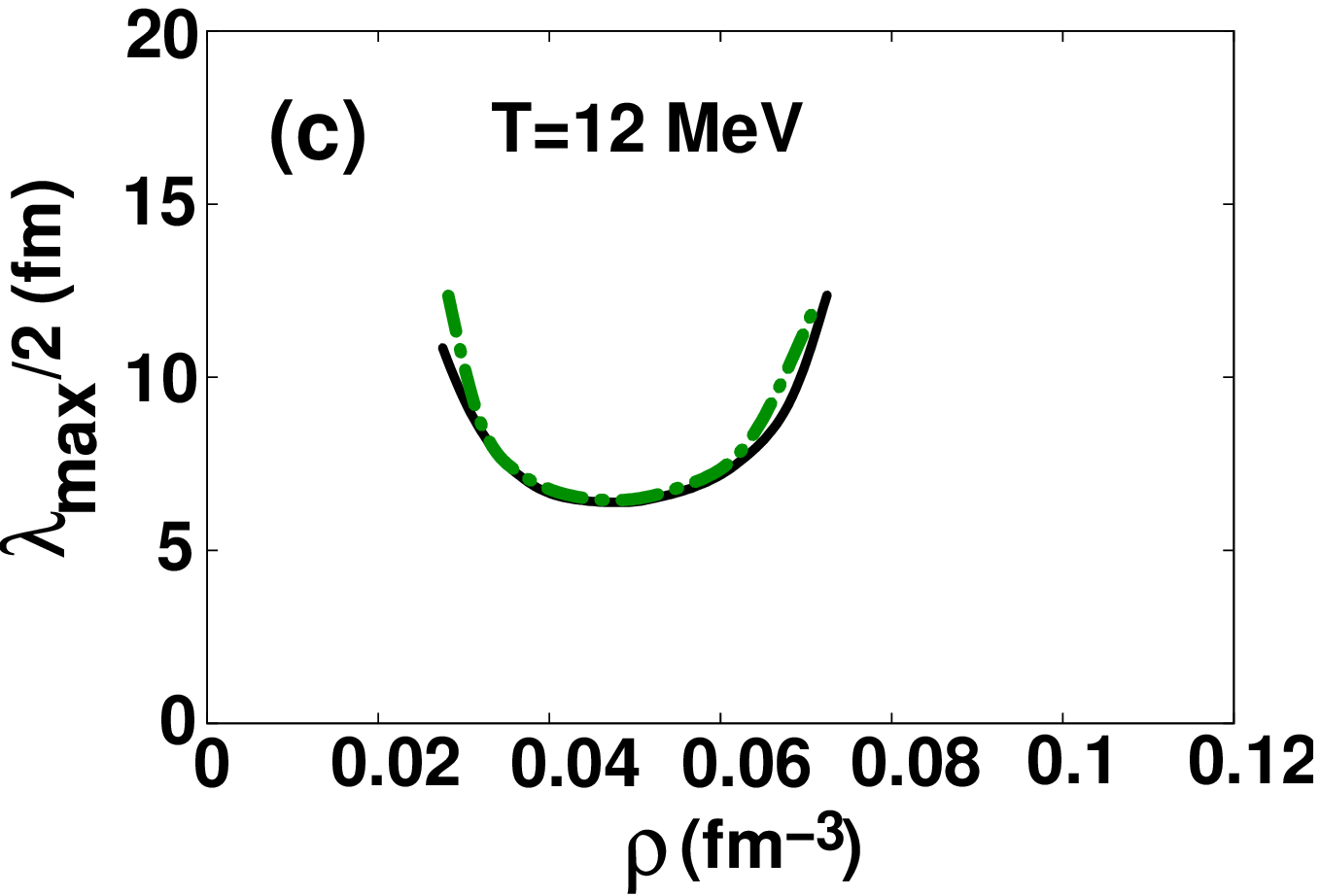}
   \end{tabular}
 \end{tabular}
\caption{(Color online) Size of the clusters in $\beta$-equilibrium with 
neutrino-trapping ($Y_L=0.4)$ at $T=5$ (a), $T=10$ (b) and $T=12$ MeV (c) for NL3 (blue, dashed), NL$\rho\delta$ (green, dot-dashed) and NL$\rho$ (black, continuous line).}%
\label{fig7}
\end{figure*}
The unstable modes of the system are calculated by replacing  the frequency
$\omega$  by  $i\Gamma$, where $\Gamma$ defines the exponential  growth rate
of the instabilities. We take the mode with the largest growth rate as the most unstable amongst all and, therefore, the one that  drives the system to the
formation of instabilities. Its half-wavelenght defines the most probable size
of the clusters (liquid) formed in the mixed (liquid-gas) phase
\cite{Lucilia-06}. In Fig. \ref{fig6} we plot the estimated size of the
clusters in $\beta$-equilibrium matter without trapped neutrinos at T=0 MeV
for the three models considered. It is shown that NL3 predicts larger clusters
as compared with the other two models, and for a smaller density range. In
Fig. \ref{fig7} we also plot the estimated size of the clusters, with trapped
neutrinos and for different temperatures: T=5, 10 and 12 MeV. When neutrinos
are included and the temperature is increased, there is a decrease in the
density range for clusterization  and an increase in the size of the
clusters. It is interesting to notice the effect of temperature on the size of
the clusters which does not agree with the conclusions of Ref. \cite{Avancini-08}
which predicts a small reduction of the cluster size with temperature (see Fig
7 of \cite{Avancini-08}). There, a simple approach was used to determine
the clusters: a zero surface thickness ansatz for the  clusters was used with a non-self consistent
surface energy. It is, therefore, important to understand this point by
describing the clusterized phase  within a  finite temperature self-consistent Thomas-Fermi  calculation.    

At T=12 MeV the NL3 parameter set predicts no clusters. Furthermore, this
model shows larger clusters and a smaller density  range of the
non-homogeneous phase as compared with the other two. At the transition
densities, NL$\rho\delta$ and NL$\rho$ present clusters with  the double of
the size as compared with the intermediate densities. For T=10 MeV, Skyrme models predict a smaller range of the instability region and also smaller clusters at the transition densities \cite{Ducoin-08}, as compared to NL$\rho\delta$ and NL$\rho$. Globally, the effect of the temperature is reducing the instability region and increasing the cluster size.

\section{Conclusions} \label{V}

In the present work we have studied the dynamical instabilities and phase
transitions in nuclear matter within the framework of the relativistic
non-linear Walecka model (NLWM) in the Mean-Field Approximation, using the
Vlasov formalism. Moreover, we have used three different parametrizations, NL3
\cite{Lalazissis-97}, NL$\rho\delta$ and NL$\rho$ \cite{Liu-02}, in order to understand the role
of $\delta$ mesons in the crust-core phase transition of a cold and a warm neutron star. 

Summarizing, the $\delta$ meson has a larger effect in neutron-rich matter, at
larger densities and lower temperatures.
For densities $\rho > 0.06$fm$^{-3}$, the $\delta$ meson increases the
distillation effect and the contrary occurs below that density.  In addition to it, the distillation effect decreases with temperature.

Besides the distillation effect we have also  tested the behaviour of the
spinodal regions in the presence of $\delta$ mesons and with the increase
of temperature. It was shown that the spinodal regions become smaller at
temperature increase, and slightly smaller with the inclusion of the $\delta$
meson. The model NL3 has a smaller envelope of the instabilities.  A larger
spinodal for NL$\rho\delta$ and NL$\rho$ is related to the finite range of the
nuclear force in this model.

In order to study the presence of a non-homogeneous phase in
$\beta$-equilibrium stellar matter we have determined the  crossing density of
the spinodals with the EoS for two cases: with neutrinos, where the constant
leptonic fraction was $Y_L=0.4$ and in $npe^-$ matter. For the first case, the
EoS crosses the spinodal for $T<11.6$ MeV, considering the NL3 parametrization
and $T<13.2$ MeV, for the NL$\rho\delta$ and NL$\rho$ parametrizations. For
the second case, the EoS crosses the spinodal region only for $T<3$ MeV. Around $T>3$ MeV, the non homogeneous phase disappears. Also, the unstable regions are larger for NL$\rho\delta$ and NL$\rho$, compared with NL3 and it was shown that the $\delta$ meson slightly decreases the instabilities.

A quantity of concern that has recently been given more attention is the
core-crust transition density. In \cite{Link-99} the authors relate this
density with the fraction of moment of inertia of compact stars. In
\cite{Centelles-09} a compilation of neutron skin thicknesses of several nuclei
brings a relation between the symmetry energy coefficients of the EoS of
nuclear matter and of nuclei described by several different models, what
allows to estimate
the core-crust transition density, which is reported to be $\sim$ 0.095
fm$^{-3}$. The values obtained in the present work  at zero temperature are
slightly lower than this value and extremely close to the findings in
\cite{Avancini-09} where a Thomas Fermi calculation for the pasta phase was
carried out.

We have determined the crust-core phase transition densities and the corresponding pressures 
as a function of temperature.
These two quantities  decrease with the temperature for NL$\rho$ and
NL$\rho\delta$ above T=2 MeV. For NL3,  the pressure slightly increases for
T$<$ 7 MeV and decreases above that temperature. The pressures are slightly larger in the presence of $\delta$ mesons. 
The inclusion of $\delta$ only slightly reduces the crossing densities and
increases the pressure at transition which is the main quantity defining the fraction of moment of inertia corresponding to the
crust \cite{Link-99}.

The estimated size of the clusters was determined as a function of temperature. NL3 shows larger clusters and a smaller range of the non-homogeneous phase as compared with the other two. At the transition densities, NL$\rho$ and NL$\rho\delta$ predict clusters with double size as
compared with an intermediate density. It was also shown that the size of the
clusters increases with temperature. The main effect of $\delta$ mesons on the size of the clusters occurs
precisely at the crust-core transition: it reduces the transition
density and the steep increase into larger clusters occurs at lower densities. A careful determination of the size of the clusters for the temperature at
which neutrino trapping occurs is important to understand whether this may
affect the interaction of the neutrinos with nuclear matter and favour the reactivation of the explosion in the dynamics of a supernova.

It has been shown that at finite temperature the behaviour of very asymmetric warm
nuclear matter depends on the model, and that $\delta$ mesons have a
stronger effect at the crust-core transition density. 
This suggests that better constraints at finite temperature on the EoS are required. 

\appendix
\section{Equations for the fields}
\label{eq_campos}
\begin{equation}
\frac{\partial^2\phi}{\partial t^2} - \nabla^2\phi +m_s^2\phi +
\frac{\kappa}{2} \phi^2 + \frac{\lambda}{6} \phi^3
= g_s\rho_s({\bf r},t) \; ,
\label{eqmphi}
\end{equation}
\begin{equation}
\frac{\partial^2\Delta_{3}}{\partial t^2} - \nabla^2\Delta_{3} +m_\delta^2\Delta_{3}=g_\delta\rho_{3s}({\bf r},t) \; ,
\end{equation}

\begin{equation}
\frac{\partial^2 V_0}{\partial t^2} - \nabla^2 V_0 + m_v^2 V_0\, =\,
g_v j_0({\bf r},t) \;,
\label{eqmv0}
\end{equation}
\begin{equation}
\frac{\partial^2 V_i}{\partial t^2} - \nabla^2 V_i + m_v^2 V_i\, =\,
g_v j_i({\bf r},t) \;,
\label{eqmv}
\end{equation}
\begin{equation}
\frac{\partial^2 b_0}{\partial t^2} - \nabla^2 b_0 + m_\rho^2 b_0\, =\,
\frac{g_\rho}{2} j_{3,0}({\bf r},t)\;,
\label{eqmb0}
\end{equation}
\begin{equation}
\frac{\partial^2 b_i}{\partial t^2} - \nabla^2 b_i + m_\rho^2 b_i\, =\,
\frac{g_\rho}{2} j_{3,i}({\bf r},t) \;,
\label{eqmb}
\end{equation}
\begin{equation}
\frac{\partial^2 A_0}{\partial t^2} - \nabla^2 A_0 \, =\,
e [j_{0p}({\bf r},t) -j_{0e}({\bf r},t)] \;,
\label{eqma0}
\end{equation}
\begin{equation}
\frac{\partial^2 A_i}{\partial t^2} - \nabla^2 A_i \, =\,
e[j_{ip}({\bf r},t) -j_{ie}({\bf r},t) ]\;,
\label{eqma}
\end{equation}
where the scalar density is
\begin{eqnarray*}
\rho_s({\bf r},t)&=&2\sum_{i=p,n}\int\frac{d^3p}{(2\pi)^3}\,
(f_{i+}({\bf r},{\bf p},t)+f_{i-}({\bf r},{\bf p},t))\,
\frac{M_i^*}{\epsilon_i} \\ \nonumber
&=&\rho_{sp}+\rho_{sn}
\end{eqnarray*}
and the isovector density is 
\begin{eqnarray*}
\rho_{3s}({\bf r},t)&=&2\sum_{i=p,n}\int\frac{d^3p}{(2\pi)^3}\,\tau_i
(f_{i+}({\bf r},{\bf p},t)+f_{i-}({\bf r},{\bf p},t))\,
\frac{M_i^*}{\epsilon_i} \qquad \\ \nonumber
&=&\rho_{sp}-\rho_{sn}.
\end{eqnarray*}
The components of the baryonic four-current density are
\begin{eqnarray*}
j_0({\bf r},t)&=&2 \sum_{i=p,n}\int\frac{d^3p}{(2\pi)^3}\,
(f_{i+}({\bf r},{\bf p},t)-f_{i-}({\bf r},{\bf p},t))\\ \nonumber 
&=& \rho_p+\rho_n \;,
\end{eqnarray*}
\begin{equation*}
{\bf j}({\bf r},t)=2\sum_{i=p,n}\int\frac{d^3p}{(2\pi)^3}\,
(f_i({\bf r},{\bf p},t)+f_{i-}({\bf r},{\bf p},t))
\frac{{\bf p}-\boldsymbol{\cal V}_i}{\epsilon_i} \;,
\end{equation*}
\begin{equation*}
j_{0e}({\bf r},t)=2 \int\frac{d^3p}{(2\pi)^3}\,
(f_{e+}({\bf r},{\bf p},t)-f_{e-}({\bf r},{\bf p},t))\; ,
\end{equation*}
\begin{equation*}
{\bf j}_e({\bf r},t)=2\int\frac{d^3p}{(2\pi)^3}\,
(f_{e+}({\bf r},{\bf p},t)+f_{e-}({\bf r},{\bf p},t)\,
\frac{{\bf p}+e\boldsymbol{\mathbf A}}{\epsilon_e} \; ,
\end{equation*}
and the components of the isovector four-current density are
\begin{eqnarray*}
j_{3,0}({\bf r},t)&=&2\sum_{i=p,n}\int\frac{d^3p}{(2\pi)^3}\, \tau_i
(f_{i+}({\bf r},{\bf p},t)-f_{i-}({\bf r},{\bf p},t))\\ \nonumber
&=&\rho_p-\rho_n \;, \\ \nonumber
{\bf j}_3({\bf r},t)&=&2\sum_{i=p,n}\int\frac{d^3p}{(2\pi)^3}\,
\frac{{\bf p}-\boldsymbol{\cal V}_i}{\epsilon_i}\\ \nonumber
&\times& \tau_i\, (f_{i+}({\bf r},{\bf p},t)+f_{i-}({\bf r},{\bf p},t)) \; ,
\end{eqnarray*}
 with
$\epsilon_i=\sqrt{({\bf p}-\boldsymbol{\cal V}_i)^2+{M_i^*}^2} \;, i=p,n \quad$
and \\
$\epsilon_e=\sqrt{({\bf p}+e{\mathbf A})^2+m_e^2} \;.$
\section{Equations of motion}
\label{eq_flut}
From the continuity equation for the density currents, we get for the components of the vector fields
\begin{eqnarray}
\omega\, \delta V_\omega^0 &=& k\, \delta V_\omega \label{contV} ,\\
\omega\, \delta b_\omega^0 &=& k\, \delta b_\omega \label{contb} ,\\
\omega\, \delta A_\omega^0 &=& k\, \delta A_\omega\,.
\label{contA}
\end{eqnarray}
Defining
$$
\delta \rho_{si}=\int \frac{d^3p}{(2 \pi)^3}\frac{p \cos \theta}
{\epsilon_{0 i}} \left[S_{\omega +}^i \frac{df_{0i+}}{dp^2} +
S_{\omega -}^i \frac{df_{0i-}}{dp^2} \right],
$$
$$
\delta \rho_i=\int \frac{d^3p}{(2 \pi)^3} p \cos \theta
\left[S_{\omega +}^i \frac{df_{0i+}}{dp^2} -
S_{\omega -}^i \frac{df_{0i-}}{dp^2} \right]
$$
with
$$ \frac{df_{0i \pm}}{dp^2} = \frac{1}{2T \epsilon_{0 i}} f_{0i \pm}
(f_{0i \pm} -1)$$
the equations of motion read
\begin{eqnarray}
i\left(\omega \mp \frac{k p \cos \theta}{\epsilon_{0 e}}\right){\cal S}_{\omega \pm}^e &=& -e\left[1 \mp \frac{\omega}{k}\frac{p \cos \theta}{\epsilon_{0 e}} \right]\delta{A}_\omega^0, \qquad  \label{Se} \\
i\left(\omega \mp \frac{k p \cos \theta}{\epsilon_{0 i}}\right )
{\cal S}_{\omega \pm}^i &=& \mp \frac{g_s\,M_i^*}{\epsilon_{0 i}}\delta\phi_\omega \nonumber \\ 
\mp \frac{g_\delta\,\tau_i\,M_i^*}{\epsilon_{0 i}}\delta\Delta_\omega
&+&\left[1 \mp \frac{\omega}{k}\frac{p \cos \theta}{\epsilon_{0 i}} \right]\delta {\cal V}_{ \omega}^{0i}  , \label{Si}\\ 
\left( \omega^2-k^2-m^2_{s,eff}\right) \delta\phi_\omega &=&
4i g_s k \sum_{i=p,n} M_i^* \delta \rho_{si} \nonumber \\ 
&+& g_s\,g_\delta\,d\rho_{3s}^0\,\delta\Delta_\omega ,  \label{fi}\\
\left( \omega^2-k^2-m^2_{\delta,eff}\right) \delta\Delta_\omega &=&
4i g_\delta k \sum_{i=p,n} \tau_i M_i^*\delta \rho_{si} \nonumber \\ 
&+& g_s\,g_\delta\,d\rho_{3s}^0\,\delta\phi_\omega  , \label{delta}\\
\left( \omega^2-k^2-m_v^2 \right)\delta V_\omega^0 &=&
4i g_v k \sum_{i=p,n} \delta \rho_i \label{V0} ,\\
\left( \omega^2-k^2-m_\rho^2 \right)\delta b_\omega^0 &=&
2i g_\rho k \sum_{i=p,n} \tau_i \delta \rho_i \label{b0} ,\\
\left( \omega^2-k^2 \right)\delta A_\omega^0 &=&
4i e k \sum_{i=p,e} (-1)^{n_i}\delta \rho_i
\label{A}
\end{eqnarray}

where 
\begin{equation*}
m^2_{s,eff}=m_s^2+\kappa\phi_0+\frac{\lambda}{2}\phi_0^2
+g_s^2 d\rho_s^0,
\label{mseff}
\end{equation*}
\begin{equation*}
m^2_{\delta,eff}=m_\delta^2+g_\delta^2 d\rho_s^0,
\label{mdeleff}
\end{equation*}
and
$$
d\rho_{3s}^0= \frac{2}{(2\pi)^3} \sum_{i=p,n}\int d^3p \,\tau_i \left(f_{0i+}+
f_{0i-}\right)\left(\frac{1}{\epsilon_{0 i}}-\frac{M_i^{*2}}{\epsilon_{0 i}^3}\right).
$$

\section{Solutions for the eigenmodes and the dispersion relation}
\label{rel_disp}

The solutions of Eqs. (\ref{contV})-(\ref{A}) form a complete set of eigenmodes that may be used to construct a general solution for an arbitrary longitudinal pertubation. Substituting the set of equations (\ref{fi})-(\ref{A}) into (\ref{Se}) and (\ref{Si}) we get a set of equations for the unknowns ${\cal S}_{\omega \pm}^i$, which lead to the following matrix equation:
\begin{equation}
\left(\begin{array}{ccccc}
a_{11}&a_{12}&a_{13}&a_{14}&a_{15}\\
a_{21}&a_{22}&a_{23}&a_{24}&0\\
a_{31}&a_{32}&a_{33}&a_{34}&a_{35}\\
a_{41}&a_{42}&a_{43}&a_{44}&0\\
0&0&a_{53}&0&a_{55}\\
\end{array}\right)
\left(\begin{array}{c}
\rho^S_{\omega p} \\
\rho^S_{\omega n} \\
\rho_{\omega p} \\
\rho_{\omega n} \\
\rho_{\omega e}
\end{array}\right)=0 \label{matriz}
\end{equation}

The amplitudes $\rho^S_{\omega i}$ (eq.(\ref{rhosi})) and $\rho_{\omega i}$ (eq.(\ref{rhoi})) are functions of the quantities ${\cal S}_{\omega \pm}^i$. The dispersion relation is obtained from the determinant of the matrix of the coefficients, namely, 
\begin{equation}
Det (a_{ij})=0 \label{det}.
\end{equation}

We define the following quantities
$$\bar \omega=\frac{\omega}{k}, \quad x= \frac{p \cos \theta}{\epsilon_{0 i}},\quad G_{\phi_i}=\frac{g_s M_i^*}{k}, \quad G_{\delta_i}=\frac{\tau_i g_\delta M_i^*}{k},$$
$$Z=\frac{1}{\left[(\bar\omega^2-\bar\omega_s^2)(\bar\omega^2-\bar\omega_\delta^2)-D_s^2\right]}, \quad D_s=\left(\frac{g_s g_\delta d\rho_{3s}^0}{k^2}\right),$$
$$ \quad \bar\omega_s^2=\frac{1}{k^2}(k^2+m^2_{s,eff}),\quad \bar\omega_\delta^2=\frac{1}{k^2}(k^2+m^2_{\delta,eff}),$$
\begin{eqnarray*}
c_s^{ij}&=&\frac{Z}{2 \pi^2 T}\left[G_{\phi_i}\left((\bar\omega^2-\bar\omega_\delta^2)G_{\phi_j}+D_s G_{\delta_j}\right)\right.\\
&+&\left. G_{\delta_i}\left((\bar\omega^2-\bar\omega_s^2)G_{\delta_j}+D_s G_{\phi_j}\right)\right]
\end{eqnarray*}
$$c_v=\frac{2}{(2 \pi)^2 T} \frac{1}{\bar \omega^2 - \bar\omega_v^2} 
\left(\frac{g_v}{k}\right)^2 (1- \bar \omega^2),
\qquad \bar\omega_v^2=\frac{1}{k^2}(k^2 + m_v^2),$$  
$$c_\rho= \frac{2}{(2 \pi)^2 T} \frac{1}{\bar \omega^2 - \bar\omega_\rho^2} 
\left(\frac{g_\rho}{2k}\right)^2 (1- \bar \omega^2), \qquad
\bar\omega_{\rho}^2=\frac{1}{k^2}(k^2+m_\rho^2),$$
$$c_e= \frac{-2}{(2 \pi)^2 T} \left(\frac{e}{k}\right)^2,$$
$$I_{\omega_\mp}(\epsilon)= \int_{-p/\epsilon}^{p/\epsilon} dx
\frac{x}{\bar \omega \pm x} = \pm \left[2 \frac{p}{\epsilon} + \bar \omega ~ln
\left| \frac{\bar \omega - p/\epsilon}{\bar \omega + p/\epsilon} \right| 
\right],$$
$$I^{ni}_{\omega_\mp}= \int_{M_i^*}^{\infty} \epsilon^n I_{\omega_\mp}(\epsilon) 
f_{0 i \mp}( f_{0 i \mp} -1) d\epsilon\, ,$$
$$A_{\omega i,\pm}^n=\int_{M_i^*}^\infty \epsilon^nd\epsilon
\int_{-p/\epsilon}^{p/\epsilon}
dx\, x\, S_{\omega \pm}^i(x,p)\, f_{0i,\pm}\left( f_{0i,\pm}-1 \right),$$
\begin{eqnarray}
\rho_{\omega i}&=&A_{\omega i +}^1-A_{\omega i -}^1, \quad i=p,n,e \label{rhoi}\\ 
\rho^S_{\omega i}&=&A_{\omega i +}^0+A_{\omega i -}^0, \quad i=p,n\, . \label{rhosi}
\end{eqnarray}

The coefficients $a_{ij}$ are defined as:
\begin{eqnarray*}
a_{11}&=&1+c_s^{pp}\left( I_{\omega  +}^{0p}- I_{\omega  -}^{0p} \right),\,\,\, 
a_{12}=c_s^{pn}\left(  I_{\omega  +}^{0p}- I_{\omega  -}^{0p} \right),\\
a_{13}&=&-(c_v+c_\rho+c_e) \left(I_{\omega  +}^{1p}+I_{\omega-}^{1p}\right),\\
a_{14}&=&-(c_v-c_\rho) \left(I_{\omega  +}^{1p}+I_{\omega -}^{1p}\right),\,\,
a_{15}= c_e \left(I_{\omega  +}^{1p}+I_{\omega -}^{1p}\right),\\
a_{21}&=&c_s^{np}\left(  I_{\omega  +}^{0n}- I_{\omega  -}^{0n} \right),\,\,
a_{22}=1+c_s^{nn}\left(  I_{\omega  +}^{0n}- I_{\omega  -}^{0n} \right),\\
a_{23}&=&-(c_v-c_\rho) \left(I_{\omega  +}^{1n}+I_{\omega-}^{1n}\right),\\
a_{24}&=&-(c_v+c_\rho) \left(I_{\omega  +}^{1n}+I_{\omega -}^{1n}\right),\,\,
a_{25}=0,\\
a_{31}&=&+c_s^{pp}\left( I_{\omega+}^{1p}+ I_{\omega  -}^{1p} \right),\,\,
a_{32}=+c_s^{pn}\left( I_{\omega+}^{1p}+ I_{\omega  -}^{1p} \right),\\
a_{33}&=&1-(c_v+c_\rho+c_e) \left(I_{\omega +}^{2p}-I_{\omega-}^{2p}\right),\\
a_{34} &=&-(c_v-c_\rho) \left(I_{\omega  +}^{2p}-I_{\omega -}^{2p}\right),\,\,
a_{35}=c_e \left(I_{\omega  +}^{2p}-I_{\omega -}^{2p}\right),\\
a_{41}&=&c_s^{np}\left(  I_{\omega  +}^{1n}+ I_{\omega  -}^{1n} \right),\,\,
a_{42}=c_s^{nn}\left(  I_{\omega  +}^{1n}+ I_{\omega  -}^{1n} \right),\\
a_{43}&=&-(c_v-c_\rho) \left(I_{\omega  +}^{2n}-I_{\omega-}^{2n}\right),\\
a_{44}&=&1-(c_v+c_\rho) \left(I_{\omega  +}^{2n}-I_{\omega -}^{2n}\right),\\
a_{51}&=&a_{52}=a_{54}=0,\\
a_{53}&=&c_e \left(I_{\omega  +}^{2e}-I_{\omega-}^{2e}\right),\,\,
a_{55}=1-c_e \left(I_{\omega  +}^{2e}-I_{\omega -}^{2e}\right).
\end{eqnarray*}

The ratios of the amplitudes are given by:
\begin{equation}
\frac{\rho_{\omega p}}{\rho_{\omega n}}=-\frac{a_{11}a_{pn}+a_{12}a_{nn}+a_{14}}{a_{11}a_{pp}+a_{12}a_{np}+a_{13}-a_{15}a_{53}/a_{55}} \label{amp1}
\end{equation}

with
$$a_{pp}=\frac{a_{22}a_{43}-a_{23}a_{42}}{a_{21}a_{42}-a_{22}a_{41}},\, \quad a_{pn}=\frac{a_{22}a_{44}-a_{24}a_{42}}{a_{21}a_{42}-a_{22}a_{41}},\,$$
$$a_{nn}=\frac{a_{44}a_{21}-a_{41}a_{24}}{a_{41}a_{22}-a_{42}a_{21}},\,  \quad a_{np}=\frac{a_{43}a_{21}-a_{41}a_{23}}{a_{41}a_{22}-a_{42}a_{21}}$$

and
\begin{equation}
\frac{\rho_{\omega e}}{\rho_{\omega p}}=-\frac{a_{53}}{a_{55}} \label{amp2}.
\end{equation}

\section*{ACKNOWLEDGMENTS}

This work was partially supported by FCT (Portugal) under grants PTDC/FIS/64707/2006, SFRH/BPD/29057/2006 and CERN/FP/83505/2008.
%

%

\end{document}